%% file: sample-aamas19.tex
\documentclass[sigconf]{aamas}  
\usepackage{balance}  
\usepackage{marvosym}

\usepackage{booktabs}

\usepackage[T1]{fontenc}
\usepackage{aecompl}
\usepackage{bm}
\usepackage{algorithm}	
\usepackage{algorithmic}
\usepackage{subfigure}
\usepackage{makecell}
\usepackage{multirow}
\usepackage{enumerate}

\setcopyright{ifaamas}  
\acmDOI{}  
\acmISBN{}  
\acmConference[AAMAS'19]{Proc.\@ of the 18th International Conference on Autonomous Agents and Multiagent Systems (AAMAS 2019)}{May 13--17, 2019}{Montreal, Canada}{N.~Agmon, M.~E.~Taylor, E.~Elkind, M.~Veloso (eds.)}  
\acmYear{2019}  
\copyrightyear{2019}  
\acmPrice{}  

\begin{document}

\title{Independent Generative Adversarial Self-Imitation Learning in Cooperative Multiagent Systems}  



\author{Xiaotian Hao*, Weixun Wang*, Jianye Hao\Letter, Yaodong Yang}
\affiliation{%
  \institution{College of Intelligence and Computing, Tianjin University}
  \city{Tianjin}
  \state{China}
  \postcode{300350}
}
\email{{xiaotianhao, wxwang, jianye.hao}@tju.edu.cn, yydapple@gmail.com}

\thanks{* Equal contribution. \Letter Corresponding author}

\begin{abstract}  
Many tasks in practice require the collaboration of multiple agents through reinforcement learning. In general, cooperative multiagent reinforcement learning algorithms can be classified into two paradigms: Joint Action Learners (JALs) and Independent Learners (ILs). In many practical applications, agents are unable to observe other agents' actions and rewards, making JALs inapplicable. In this work, we focus on independent learning paradigm in which each agent makes decisions based on its local observations only. However, learning is challenging in independent settings due to the local viewpoints of all agents, which perceive the world as a non-stationary environment due to the concurrently exploring teammates. In this paper, we propose a novel framework called Independent Generative Adversarial Self-Imitation Learning (\textbf{IGASIL}) to address the coordination problems in fully cooperative multiagent environments. To the best of our knowledge, we are the first to combine self-imitation learning with generative adversarial imitation learning (GAIL) and apply it to cooperative multiagent systems. Besides, we put forward a Sub-Curriculum Experience Replay mechanism to pick out the past beneficial experiences as much as possible and accelerate the self-imitation learning process. Evaluations conducted in the testbed of StarCraft unit micromanagement and a commonly adopted benchmark show that our \textbf{IGASIL} produces state-of-the-art results and even outperforms JALs in terms of both convergence speed and final performance.
\end{abstract}

%


\maketitle


\input{samplebody-conf}


\bibliographystyle{ACM-Reference-Format}  
\balance  
\bibliography{sample-bibliography}  

\end{document}

%% file: samplebody-conf.tex
\section{Introduction}

With the advance of deep neural network \cite{lecun2015deep, goodfellow2016deep}, Deep Reinforcement Learning (DRL) approaches have made significant progress for a number of applications including Atari games \cite{mnih2015human}, Go \cite{silver2016mastering}, game theory \cite{leibo2017multi,wang2018towards} and robot locomotion and manipulation \cite{levine2016end,schulman2015high}.
In practice, a large number of important applications can be naturally modeled as cooperative multiagent systems. Examples include coordination of robot swarms \cite{huttenrauch2017guided}, coordination of autonomous vehicles \cite{cao2013overview}, network packet delivery \cite{ye2015multi}, managing air traffic flow \cite{agogino2012multiagent} and energy distribution \cite{yang2018recurrent}.

However, directly applying single-agent reinforcement learning approaches such as Q-learning to cooperative multiagent environments behaves poorly. Thus, effective coordination mechanism needs to be incorporated into agents' learning strategies to address the cooperative multiagent problems. 
 Multiagent reinforcement learning (MARL) can be generally classified in to two paradigms \cite{claus1998dynamics}: Joint Action Learners (JALs) and Independent Learners (ILs). JALs observe the rewards and actions (policies) taken by all agents whose information is explicitly considered during policy update, whereas ILs make decisions based on their local observations, actions and rewards only. Under the JAL paradigm, MADDPG \cite{lowe2017multi} is a recently proposed approach for multiagent games with large continuous state space and action space.
By taking the other agents' observations and policies directly into consideration, MADDPG learns a centralized critic and uses the centralized critic to provide a better guidance for the policy update. However, MADDPG does not 
consider some specially designed mechanisms for handling the multiagent cooperative challenges when dealing with difficult cooperation environments (e.g., sparse rewards, high miss-coordination penalties,  exploration-exploitation trade-off \cite{matignon2012independent}).
Besides, due to the inaccessibility of all other agents' states and actions in practice and the exponential growth of the state-action space in the number of agents, JALs are difficult to be applied to practical applications.

Avoiding the above two restrictions, ILs are more universally applicable and have been widely studied over the past years, e.g., Distributed Q-learning \cite{lauer2000algorithm}, Hysteretic Q-learning \cite{matignon2007hysteretic} and Lenient Learners \cite{panait2006lenient}. However, for ILs, one typical issue is that each agent's policy is changing as training progresses, and the environment becomes non-stationary from the perspective of any individual agent since other agents' policies are changing concurrently. Hysteretic Q-learning \cite{matignon2007hysteretic} and Lenient Learners \cite{panait2006lenient} are proposed to facilitate multiple reinforcement learning agents to overcome the independent learning problems (e.g., the non-stationary problem \cite{matignon2012independent}). 
Very recently, the idea of hysteretic Q-learning and lenient learners has been successfully applied 
to deep multiagent reinforcement learning settings \cite{omidshafiei2017deep,palmer2018lenient}. 
However, all these approaches are Q-learning based methods and are naturally suitable for settings with discrete action space only. Therefore, it's difficult to apply these approaches to solve the cooperative multiagent continuous control tasks.

In this work, we propose a novel framework under the independent learning paradigm called independent generative adversarial self imitation learning (IGASIL), which conducts both learning and execution in a fully decentralized manner. In the framework, there are $n$ independent agents cooperatively solving a task without knowing other agents' policies and making decisions based on their own local observations.
Initially, each agent maintains a positive buffer and a normal buffer. At run time, each agent interacts with the environment independently according to the current policy. The resulting trajectory is stored twice in the positive buffer and the normal buffer. The positive buffer is a specially designed sub-curriculum experience replay which continuously helps to pick out and reserve preferable experiences the agent has experienced.
Combining self-imitation learning with generative adversarial imitation learning, each agent trains a discriminator using samples from these two buffers whose target is to capture the features of the past good experiences. Besides the environment rewards, each agent receives additional rewards from the discriminator, which would guide the agents to imitate from the past good experiences and do more exploration around these high-reward regions. Once the agents find better policies, they will produce higher quality trajectories. Thus, the learning will turn into a virtuous circle until a good cooperation is achieved.

The main contributions of this paper can be summarized as follows.

\begin{enumerate}[(1)]
\vspace{-0.5em}
  \item To the best of our knowledge, we are the first to combine self imitation learning with generative adversarial imitation learning and propose a novel framework called Independent Generative Adversarial Self Imitation Learning (IGASIL) to address the multiagent coordination problems. 
  \item We put forward a Sub-Curriculum Experience Replay mechanism to accelerate the self-imitation learning process.
  \item IGASIL is well applicable to both discrete and continuous action spaces and can be integrated with any Policy Gradient or Actor-Critic algorithm in fully cooperative multiagent environments.
  \item Besides, our proposed method follows the decentralized training pattern which does not require any communication among agents during learning.  
  \item Experimental results show that our method outperforms state-of-the-art in cooperative multiagent continuous and discrete control tasks in terms of both convergence speed and final performance.
\end{enumerate}

\section{Background} \label{Section:Background}
\subsection{Markov Decision Process}
We use the tuple $(S,A,P,r,\rho_0,\gamma)$ to define an infinite-horizon, discounted Markov decision process (MDP), where $S$ represents the state space, $A$ represents the action space, $P:S\times A\times S \rightarrow [0,1]$ denotes the transition probability distribution, $r:S \times A \rightarrow \mathbb{R}$ denotes the reward function, $\rho_0 \rightarrow [0,1]$ is the distribution of the initial state $s_0$, and $\gamma \in (0,1)$ is the discount factor. Let $\pi_\theta$ denote a stochastic policy $\pi:S\times A\rightarrow[0,1]$, where $\theta$ is the parameter of the policy. The performance of a stochastic policy $\pi_\theta$ is usually evaluated by its expected cumulative discounted reward $J_{\pi_\theta}$:
\begin{equation}\label{Equation:Return}
J_{\pi_\theta}={\mathbb{E}}_{\rho_0,P,\pi_\theta}[\sum_{t=0}^{\infty}\gamma^t r(s_t,a_t)]
\end{equation}
Reinforcement Learning (RL) \cite{sutton1998reinforcement} is a set of algorithms trying to infer a policy $\pi_{\theta}$, which maximizes the expected cumulative discounted reward $J_{\pi_{\theta}}$ when given access to a reward signal $r(s,a)$. 

\subsection{Generative Adversarial Imitation Learning}
Imitation learning is also known as learning from demonstrations or apprenticeship learning, whose goal is to learn how to perform a task directly from expert demonstrations, without any access to the reward signal $r(s,a)$. Recent main lines of researches within imitation learning are behavioural cloning (BC) \cite{pomerleau1991efficient,bojarski2016end}, which performs supervised learning from observations to actions when given a number of expert demonstrations; inverse reinforcement learning (IRL)\cite{abbeel2004apprenticeship}, where a reward function is estimated that explains the demonstrations as (near) optimal
behavior; and generative adversarial imitation learning (GAIL) \cite{ho2016generative,song2018multi,behbahani2018learning,bhattacharyya2018multi}, which is inspired by the generative adversarial networks (GAN) \cite{goodfellow2014generative}. 
Let $T_E$ denote the trajectories generated by the behind expert policy $\pi_E$, each of which consists of a sequence of state-action pairs. In the GAIL framework, an agent mimics the behavior of the expert policy $\pi_E$ by matching the generated state-action distribution $\rho_{\pi_\theta}(s,a)$ with the expert's distribution $\rho_{\pi_E}(s,a)$. The state-action visitation distribution (occupancy measure \cite{ho2016generative}) of a policy $\pi_\theta$ is defined as:
\begin{equation}
\label{Equation:occupancy_measure}
  \rho_{\pi_\theta}(s,a)=\pi_\theta(a|s)\sum_{t=0}^{\infty}\gamma^tp(s_t=s|\pi_\theta)
\end{equation}
where $p(s_t=s|\pi_\theta)$ is the probability of being in state $s$ at time $t$ when starting at state $s_0\sim\rho_0$ and following policy $\pi_\theta$. Thus, $J_\theta$ can be written as:

\begin{equation}
\label{Equation:occupancy_measure}
\begin{aligned}
  J_{\pi_\theta}&={\mathbb{E}}_{\rho_0,P,\pi_\theta}[\sum_{t=0}^{\infty}\gamma^t r(s_t,a_t)] \\
  &=\sum_{t=0}^{\infty}\sum_{s}p(s_t=s|\pi_\theta)\sum_{a}\pi_\theta(a|s)\gamma^t r(s_t,a_t) \\
  &=\sum_{s}\sum_{a}\pi_\theta(a|s)\sum_{t=0}^{\infty}p(s_t=s|\pi_\theta)\gamma^t r(s_t,a_t) \\
  &=\sum_{s,a}\rho_{\pi_\theta}(s,a)r(s_t,a_t)
\end{aligned}
\end{equation}
which only depends on the discounted state-action visitation distribution $\rho_{\pi_{\theta}}(s,a)$. The optimum is achieved when the distance between these two distributions is minimized as measured by Jensen-Shannon divergence. The formal GAIL objective is denoted as:

\begin{equation}
\label{Equation:GAIL_target}
\begin{aligned}
  &\min_\theta \max_w \mathbb{E}_{(s,a) \sim \rho_{\pi_E}(s,a)}[\log(\emph{D}_w(s,a))] + \\
  &\mathbb{E}_{(s,a) \sim \rho_{\pi_\theta}(s,a)}[\log(1-\emph{D}_w(s,a))] - \lambda_H\emph{H}(\pi_\theta)
\end{aligned}
\end{equation}
where $\emph{D}_w$ is a discriminative binary classifier parameterized by $w$ which tries to distinguish state-action pairs from the trajectories generated by $\pi_\theta$ and $\pi_E$,
$H(\pi_\theta) \triangleq \mathbb{E}_{\rho_0,P,\pi_\theta}[\sum_{t=0}^{\infty}\gamma^t(-\log \pi_\theta(a|s))]$
is the $\gamma-$discounted causal entropy of policy $\pi_\theta$ \cite{bloem2014infinite} and $\lambda_H$ is the coefficient. Unlike GANs, the original GAIL requires interactions with the environment/simulator to generate state-action pairs, and thus the objective (\ref{Equation:GAIL_target}) is not differentiable end-to-end with respect to the policy parameter $\theta$. Hence, optimization of the policy requires RL techniques based on Monte-Carlo estimation of policy gradients. The optimization over the GAIL objective is performed by alternating between $K$ gradient step to increase (\ref{Equation:GAIL_target}) with respect to the discriminator parameters $w$, and a Trust Region Policy Optimization (TRPO) step to decrease (\ref{Equation:GAIL_target}) with respect to the policy parameters $\theta$ (using $\log(\emph{D}_w(s,a))$ as the reward function).

\subsection{Sample-Efficient GAIL}
One of the most important advantages of GAIL is that it can obtain a higher performance than behavioral cloning when given only a small number of expert demonstrations. However, a large number of policy interactions with the learning environment are required for policy convergence. As illustrated in \cite{2018arXiv180902925K}, while GAIL requires as little as 200 expert frame transitions 
to learn a robust reward function on most MuJoCo \cite{todorov2012mujoco} tasks, the number of policy frame transitions sampled from the environment can be as high as 25 million in order to reach convergence, which is  intractable for real-world applications.
To this end, \cite{2018arXiv180902925K} address the sample inefficiency issue via incorporating an off-policy RL algorithm 
and an off-policy discriminator to dramatically decrease the sample complexity by many orders of magnitude. Experimental results show that their off-policy approach works well even without using the importance sampling. 

\subsection{Self-Imitation Learning}
In an environment with the very sparse reward, it's difficult to learn the whole task at once. It is natural to master some basic skills for solving easier tasks firstly. e.g., In Montezuma's Revenge (an Atari game), the agent needs to pick up the key and then open the door. Directly learning opening the door is hard due to the poor exploration, but it is easier to master picking up the key at first. Based on this idea, self-imitation learning (SIL) \cite{oh2018self} is a very recent approach proposed to solve the sparse reward problem by learning to imitate the agent's own past good experiences. In brief, SIL stores previous experiences in a replay buffer and learns to imitate the experiences when the return is greater than the agent's expectation. Experimental results show that this bootstrapping approach (learn to imitate the agent's own past good decisions) is highly promising on hard exploration tasks 
A proper level of exploitation of past good experiences during learning can lead to a deeper exploration (moving to the deeper region) of the learning environment. Similar idea and results can also be found in \cite{kang2018policy}. 


\section{Problem Description}
The setting we are considering is a fully cooperative partially observable Markov game \cite{littman1994markov}, which is a multiagent extension of a Markov decision process (MDPs). A Markov game for $N$ agents is defined by a set of states $S$ describing the possible configurations of all agents and environment, a set of actions $A_1, ..., A_N$ and a set of observations $O_1, ..., O_N$ for each agent. Initial states are determined by a distribution $\rho_0: S \to [0, 1]$. State transitions are determined by a function $P: S \times A_1 \times ... \times A_N \times S \to [0,1]$. For each agent $i$, rewards are given by function $r_i: S \times A_1 \times ... \times A_N \to \mathbb{R}$, observations are given by function $o_i: S \to O_i$. To choose actions, each agent $i$ uses a stochastic policy $\pi_i: O_i \times A_i\to [0,1]$. The joint policy $\bm{\pi}$ of all agents is defined as $\bm{\pi}:$ $\langle \pi_1 , ... , \pi_N \rangle$. The joint action is represented as $\bm{a}=\langle a_1,...,a_n\rangle$. 
We consider a finite horizon setting, with episode length $T$. 
If all agents receive the same rewards ($r_1=r_2=...=r_N$), the Markov game is fully cooperative, which means a best-interest action of one agent is also a best-interest action of all agents. Besides, we only consider the environments with deterministic reward functions at present.

In the following of this paper, we are going to analyze and deal with the coordination problems under the following two difficult cooperative environments: (1) cooperative endangered wildlife rescue; (2) decentralised StarCraft micromanagement from an independent perspective. For example, in the cooperative endangered wildlife rescue task, there are $N$ slower independent rescue agents which have to cooperatively chase and rescue one of the $M$ faster wounded animals in a randomly generated environment with continuous state and action spaces. Each rescue agent makes decisions (go north, south, east, or west to chase one of the $M$ animals) based on its local observation only and can't observe the others' policies \textbf{(local observation and continuous action space)}. Only when the $N$ rescue agents chase and capture the same wounded animal simultaneously, will they get a reward based on the caught animal's value. So, the reward is very sparse and it's hard for the independent rescuers to explore \textbf{(sparse rewards)}. Besides, the changing of the other agents' policies (e.g. the other agents' move to different directions for exploration instead of cooperatively capturing the same animal with the current agent) will influence the reward of current agent's action. 
Therefore, the environment becomes non-stationary from the perspective of each individual rescuer \textbf{(non-stationary)}. If a rescue agent changes its actions too quickly when perceiving the changed reward (due to the others' explorations), the others will change their policies in their turn \textbf{(exploration-exploitation)}. Moreover, different wounded animals has different rewards and different penalties (the animal with the higher reward also has the higher penalty for miss-coordination). So, to avoid punishment, the rescuers prefer to capture the animals with the lowest reward instead of the global optimal one \textbf{(high penalty and shadowed equilibrium)}. Detail settings of the game are shown in Section \ref{exp:effectiveness:predator-prey}. Thus, it's hard to coordinate the independent learners to achieve successful cooperation and converge to a better equilibrium. Thus, additional cooperation mechanisms are needed.

\section{Independent Self-Imitation Learning Framework} \label{Section:IGASIL_Framework}

\begin{figure}[!htb]
\centering
{\includegraphics[height=1.5in,width=3.0in,angle=0]{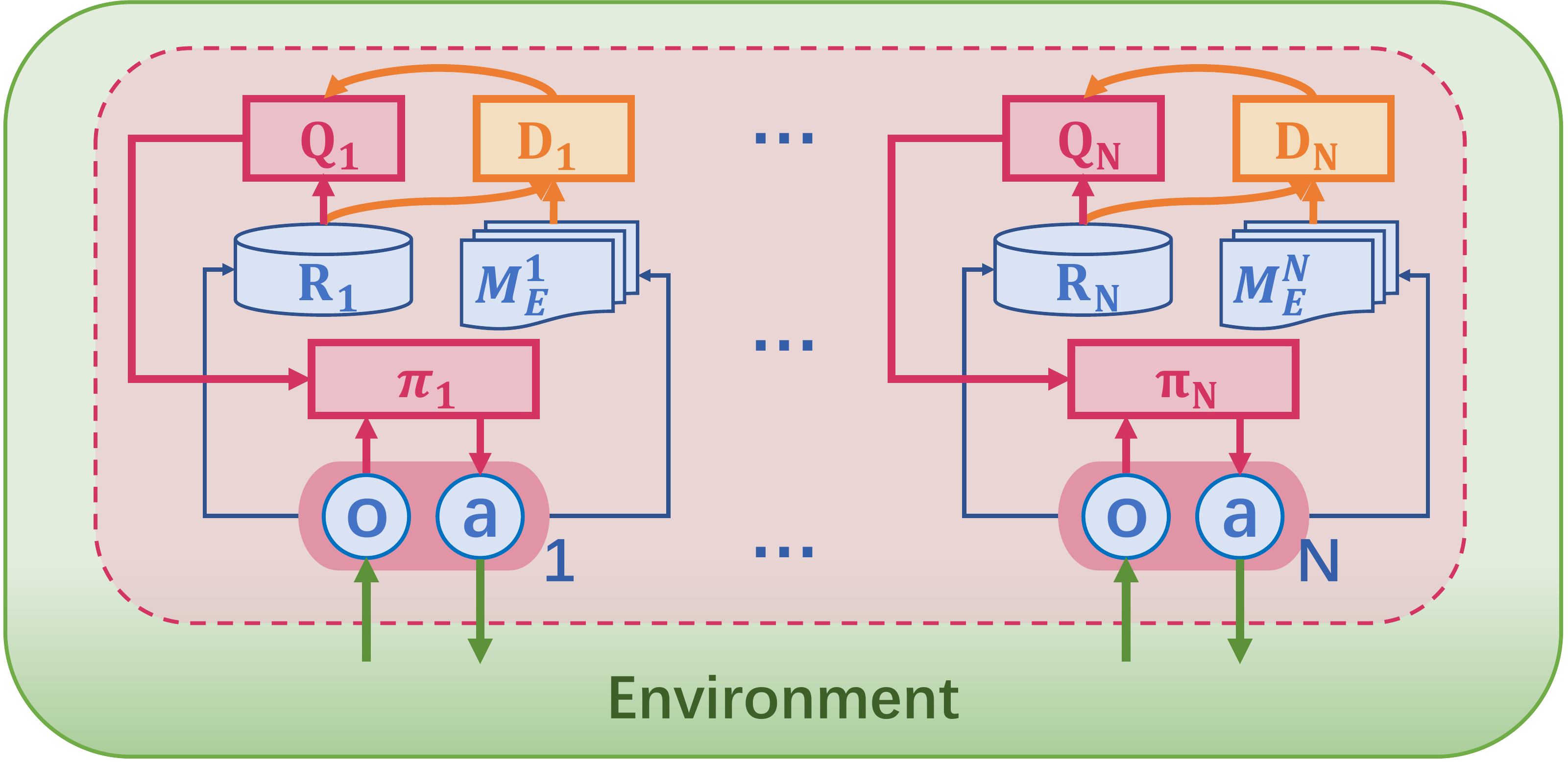}}
\caption{IGASIL framework for cooperative multiagent systems.}
\label{figure:framework}
\end{figure}

\subsection{Independent Generative Adversarial Self-Imitation Learning} \label{SubSection:method}

Combining the idea of self-imitation learning with GAIL, we propose a novel independent generative adversarial self-imitation learning (IGASIL) framework aiming at facilitating the coordination procedure of the interactive agents, reducing the learning variance and improving the sample efficiency. The learning procedure for each independent learner $i$ is summarized in Algorithm \ref{Algorithm:IGSIL}.

\begin{algorithm}[h]
    \caption{Independent Generative Adversarial Self-Imitation Learning}
    \label{Algorithm:IGSIL}
    \begin{algorithmic}[1]
        \STATE \textbf{Input:} For each agent $i$, initial parameters of actor, critic, discriminator $\theta_i^0, \phi_i^0, w_i^0$, sub-curriculum experience replay $M_E^i$ and a normal replay buffer $R_i$.
        \STATE Initialize $M_E^i \leftarrow \varnothing$, $R_i \leftarrow \varnothing$.
        \FOR{n=0,1,2,...}
            \STATE Sample a trajectory $T_n^i \sim \pi_{\theta_i}$.
            \STATE Store $T_n^i$ in $R_i$.
            \STATE Store $T_n^i$ in $M_E^i$ according to Algorithm \ref{Algorithm:PHER}.
            \STATE Sample state-action pairs $X_n \sim R_i$ and $X_E \sim M_E$ with the same batch size.
            \STATE Update $w_i^n$ to $w_i^{n+1}$ by ascending with gradients:
            \begin{equation}
            \setlength{\abovedisplayskip}{4pt}
            \setlength{\belowdisplayskip}{-1pt}
            \begin{split}
                \Delta_{w_i^n}= & \mathbb{\hat{E}}_{X_E}[\log(\emph{D}_{w_i^n}(s,a))] + \\ & \mathbb{\hat{E}}_{X_n}[\log(1-\emph{D}_{w_i^n}(s,a))]
            \end{split}
            \end{equation}
            \STATE Sample (s,a,r,s',done) tuples $X_n^\prime \sim R_i$.
            \STATE Calculate the imitation reward for each $(s,a) \sim X_n^\prime$ by:
            \begin{equation}
            \setlength{\abovedisplayskip}{4pt}
            \setlength{\belowdisplayskip}{-4pt}
            \begin{aligned}
              r_{imit}(s,a) = \log(\emph{D}_{w_i^n}(s,a)) - \log(1-\emph{D}_{w_i^n}(s,a))
            \end{aligned}
            \end{equation}
            \STATE Calculate the reshaped reward for each $(s,a) \sim X_n^\prime$ by:
            \begin{equation}
            \label{reward shaping}
            \setlength{\abovedisplayskip}{4pt}
            \setlength{\belowdisplayskip}{-4pt}
            \begin{aligned}
              r^\prime(s,a) = r + \lambda_{imit} * r_{imit}(s,a)
            \end{aligned}
            \end{equation}
            \STATE Replace $r$ in $X_n^\prime$ with $r^\prime(s,a)$.
            \STATE Using samples $X_n^\prime$ to update the policy parameter $\theta_i^n$ to $\theta_i^{n+1}$ and the critic parameter $\phi_i^n$ to $\phi_i^{n+1}$ according to DDPG \cite{lillicrap2015continuous} (off-policy A2C \cite{degris2012off})
        \ENDFOR
    \end{algorithmic}
\end{algorithm}

An illustration of our IGASIL is shown in Figure (\ref{figure:framework}). Initially, each agent $i$ maintains a sub-curriculum experience replay buffer $M_E^i$ and a normal buffer $R^i$ (Algorithm \ref{Algorithm:IGSIL}, Line 1-2). At run time, each agent interacts with the environment independently according to the current policy. The resulting trajectory is stored in $M_E^i$ and $R^i$ respectively (stored twice) (Algorithm \ref{Algorithm:IGSIL}, Line 4-6). The normal buffer $R^i$ is used for off-policy training, which will be discussed in Section \ref{Subsection:Sample-efficient GASIL}. The sub-curriculum experience replay buffer $M_E^i$ of each agent preserves the past useful skills (demonstrations) for future use, which will be detailed in Section \ref{SubSection:replay buffer}. We consider these useful demonstrations in $M_E^i$ as self generated expert data and regard the policy behind these self-generated demonstrations as $\pi_E^i$ for each agent $i$. At the same time, each agent trains a discriminator $D_i$ using samples from these two buffers whose target is to capture the features of the past good experiences. (Algorithm \ref{Algorithm:IGSIL}, Line 7-8).
Then, each agent begins to update its policy based on two types of rewards: (1) the imitation rewards given by the discriminator $D_i$ (Algorithm \ref{Algorithm:IGSIL}, Line 9-10), which will be discussed in Section \ref{Subsection:Unbiased imitation reward}; (2) the original environment rewards, which are combined according to Equation (\ref{reward shaping}), in which $\lambda_{imit}$ control the weight of the imitation reward\footnote{In our settings, we grow the $\lambda_{imit}$ exponentially as learning progresses (One intuition is that as the training progresses, the trajectories produced by the agent becomes better and better. Thus, the agent should pay more attention to these better ones).} (Algorithm \ref{Algorithm:IGSIL}, Line 11-12).
The final reshaped reward $r^\prime(s,a)$ will encourage each agent to explore more around the nearby region of the past good experience to check whether a better coordination can be achieved or have been achieved.
After that, the policy of each agent is updated according to the corresponding update rules (Algorithm \ref{Algorithm:IGSIL}, Line 13). Thus, under the guidance of the discriminator, the past good experiences and skills are dynamically reused. After that, each agent's policy is more likely updated towards a better direction independently. 
Though we use the off-policy actor-critic approaches (DDPG and off-policy A2C) in our algorithm, our self-imitation framework can be integrated with any policy gradient or actor-critic methods.


Since all independent agents receive exactly the same reward and use the same learning approach (same parameters and settings), the positive trajectories stored in $M_E^i$ would be stored in a synchronized way, which means the agents could cooperatively imitate the "same" past good experience in a distributed way. Thus, all independent agents would have the same behavioral intentions (e.g., jointly imitating the same past good experience and doing deeper exploration, which we call "the joint intention") during learning. As the "joint" imitation learning progresses, the policy of each agent would be induced to update towards the "same" direction. Consequently, the non-stationary and learning issues can be alleviated. 

\subsubsection{\textbf{Sample-efficient GASIL}} \label{Subsection:Sample-efficient GASIL}
One limitation of GAIL is that it requires a significant number of interactions with the learning environment in order to imitate an expert policy \cite{2018arXiv180902925K}, which is also the case of our settings. To address the sample inefficiency of GASIL, we use off-policy RL algorithms (Here, we use DDPG and off-policy A2C) and perform off-policy training of the GAIL discriminator performed in such way: for each agent $i$, instead of sampling trajectories from the current policy directly, we sample transitions from the replay buffer $R_i$ collected while performing off-policy training:

\begin{equation}
\label{Equation:GAIL off_target}
\begin{aligned}
  &\min_\theta \max_w \mathbb{\hat{E}}_{(s,a) \sim \pi_E^i}[\log(\emph{D}_{w_i}(s,a))] + \\
  &\mathbb{\hat{E}}_{(s,a) \sim R_i}[\log(1-\emph{D}_{w_i}(s,a))] - \lambda_H\emph{H}(\pi_\theta^i)
\end{aligned}
\end{equation}
Equation (\ref{Equation:GAIL off_target}) tries to match the occupancy measures between the expert and the distribution induced by the replay buffer $R_i$ instead of the latest policy $\pi_i$.
It has been found that the off-policy GAIL works well in practice even without using importance sampling \cite{2018arXiv180902925K}. As will be shown in Section \ref{exp:sample-efficient}, we also observe similar phenomenons in our cooperative endangered wildlife rescue environment.

\subsubsection{\textbf{Unbiased imitation reward}} \label{Subsection:Unbiased imitation reward}
Another problem of GAIL is that either $r_{imit}(s,a)=-log(1-D(s,a))$ or $r_{imit}(s,a)=log(D(s,a))$ (which is often used as the reward function in GAIL approaches) has reward biases that can either implicitly impose prior knowledge about the true reward, or alternatively, prevent the policy from imitating the optimal expert \cite{2018arXiv180902925K}. We summarize the reason of the two rewards' bias here: (1) $-log(1-D(s,a))$ is always positive and potentially provides a survival bonus which drives the agent to survive longer in the environment to collect more rewards. (2) $log(D(s,a))$ is always negative and provides a per step penalty which drives the agent to exit from the environment earlier. Thus, to stabilize the training process of our IGASIL, we use a more stable reward function as shown in Equation (\ref{Equation:stable imitation reward}). Similar analysis can be found in \cite{fu2017learning} and \cite{2018arXiv180902925K}.

\begin{equation}
\label{Equation:stable imitation reward}
\begin{aligned}
  r_{imit}(s,a) = log(D(s,a)) - log(1-D(s,a))
\end{aligned}
\end{equation}

\subsection{Sub-Curriculum Experience Replay} \label{SubSection:replay buffer}
In a complex cooperative game, a series of actions need to be taken simultaneously by all agents to achieve a successful cooperation. However, due to the independent learning agents (ILs) interacting with the environment according to their own observations and policies without any communication, each agent might randomly take different actions for exploration at the same state. But, to achieve a perfect cooperation, each agent must exactly select the "right" action at all states. This means the collected trajectories of successful cooperation are very few during learning. So, it's difficult for the independent agents to grasp all these series of actions simultaneously to achieve perfect cooperation at once. Therefore, additional mechanisms are needed to induce the individual agents to gradually pick the "right" actions at the same state.

Curriculum learning is an extension of transfer learning, where the goal is to automatically design and choose a sequence of tasks (i.e. a curriculum) $T_1, T_2, ... T_t$ for an agent to train on, such that the learning speed or performance on a target task $T_t$ will be improved \cite{narvekar2017autonomous}. Inspired by this idea, we want our independent learning agents to follow the curriculum learning paradigm. For example, it's easier for the agents to firstly learn to cooperate at some easier states. And then, reusing the basic skills learned in the previous step, the agents would gradually achieve deeper cooperation and finally are able to solve the target task. The main idea is that the past useful skills can be reused to facilitate the coordination procedure. Similar ideas have been applied to a series of curriculum learning tasks \cite{narvekar2016curriculum}, \cite{narvekar2016source}. In our settings, we consider a whole trajectory as an instance of solving the target task. Our goal is to find and reuse the past useful skills/experiences for each agent to accelerate the cooperation process. An intuitive way is to pick out these useful experiences by rewards. One example is that given two trajectories with rewards $[0, +1, +3, +1, 0, 0, -20]$ and $[0, 0, 0, 0, 0, 0, -15]$, though the total rewards are both $-15$ (low), there is still some useful experience included in the first trajectory (e.g.: the sub-trajectory $[0, +1, +3, +1]$ with a total reward +5 still demonstrates some good behaviors). By imitating the behaviors from these good sub-trajectories, the agents can still grasp some useful cooperation skills. 
Another example is considering a trajectory with sparse rewards $[0, 0, 0, ..., 0, 0, +1]$ (only receiving +1 at the terminal state), imitating from the sub-trajectories near the terminal state (e.g. $[0, +1]$, $[0, 0, +1]$) would drive the agent to quickly master skills around the terminal state, reduce unnecessary explorations, and thus ease the reward backpropagation problem when rewards are sparse, which is similar to the idea of reverse curriculum generation for reinforcement learning \cite{florensa2017reverse}.

\begin{algorithm}[h]
    \caption{Sub-Curriculum Experience Replay}
    \label{Algorithm:PHER}
    \begin{algorithmic}[1]
        \STATE \textbf{Given:} A learning policy $\pi_{\theta_i}$ for agent $i$.
        \STATE Initialize the min-heap based positive trajectory buffer $M_E^i$.
        \FOR{n=0,1,2,...}
            \STATE Sample a trajectory $T_n^i \sim \pi_{\theta_i}$.
            \STATE Calculate the discounted Return of $T_n^i$ as $R_{T_n^i}$.
            \STATE Store $(T_n^i, R_{T_n^i})$ in $M_E^i$. \qquad \qquad \qquad $\rhd$ \emph{resorting by priority}
            \FOR{j=0,N}
                \STATE Randomly sample a sub-trajectory $sub_j(T_n^i)$ from $T_n^i$ without repetition.
                \STATE Calculate the discounted Return of $sub_j(T_n^i)$ as $R_{sub_j(T_n^i)}$.
                \STATE Store $(sub_j(T_n^i), R_{sub_j(T_n^i)})$ in $M_E^i$.  \ \, $\rhd$ \emph{resorting by priority}
            \ENDFOR
            \STATE Update the policy $\pi_{\theta_i}$ according to Algorithm \ref{Algorithm:IGSIL}.
        \ENDFOR
    \end{algorithmic}
\end{algorithm}

Given the above analysis, we build a min-heap based sub-curriculum experience replay (SCER) $M_E^i$ with a small buffer size $k$ for each agent $i$ to continuously maintain the past beneficial experiences. The formal description of our SCER is summarized in Algorithm \ref{Algorithm:PHER}. The priority of the min-heap is based on the return of each trajectory/sub-trajectory. To pick out the beneficial (useful) experiences as much as possible, we randomly sample some sub-trajectories from a given trajectory without repetition (Algorithm \ref{Algorithm:PHER}, Line 8), calculate their discounted returns and feed them into $M_E^i$ (Algorithm \ref{Algorithm:PHER}, Line 9-10). The ranking and filtering of the trajectories are processed within $M_E^i$. Since $M_E^i$ is built based on a min-heap, it can be viewed as a filter which keeps the latest top-k-return trajectories/sub-trajectories (Algorithm \ref{Algorithm:PHER}, Line 10). 
The positive buffer $M_E^i$ usually starts by storing suboptimal trajectories (e.g., killing only few enmies in StarCraft micromanagement games), and our sub-curriculum ER with IGASIL allows each agent to learn better sub-policies in the subspaces. Based on the pre-learned skills, the agents are easier to explore to the deeper regions and the positive buffer will receive trajectories with higher quality. This leads to agents learning better coordinated policies in return.

\section{Experiments}
In the following experiments, we evaluate the effectiveness of our IGASIL framework in two cooperative multiagent benchmarks: (1) cooperative endangered wildlife rescue, which has the very sparse reward and high miss-coordination penalty \cite{lowe2017multi}; (2) decentralised StarCraft micromanagement, which has multi types of units, strong stochasticity and uncertainty \cite{foerster2017counterfactual,peng2017multiagent,rashid2018qmix}.

\subsection{Effectiveness of our Approach} \label{exp:effectiveness}

\textbf{Architecture \& Training.}
In this paper, all of our policies, critics and discriminators are parameterized by a two-layer ReLU MLP (Multilayer Perceptron) followed by a fully connected layer activated by tanh functions for DDPG's policy nets(DDPG is used only for the animal rescue game), softmax functions for AC's policy nets and sigmoid functions for all discriminators. Only in decentralised StarCraft micromanagement task 
, we share the parameters among the homogeneous agents (units with the same type) to accelerate the training process. 
The code has been published on GitHub \footnote{\url{https://github.com/tjuHaoXiaotian/GASIL}}.

\subsubsection{Cooperative Endangered Wildlife Rescue} \label{exp:effectiveness:predator-prey}

\

\begin{figure}[!htb]
\vspace{-0.25cm}  %
\setlength{\abovecaptionskip}{0.2cm}   
\setlength{\belowcaptionskip}{-0.3cm}   
\centering
\subfigure[] {\includegraphics[height=0.8in,width=0.8in,angle=0]{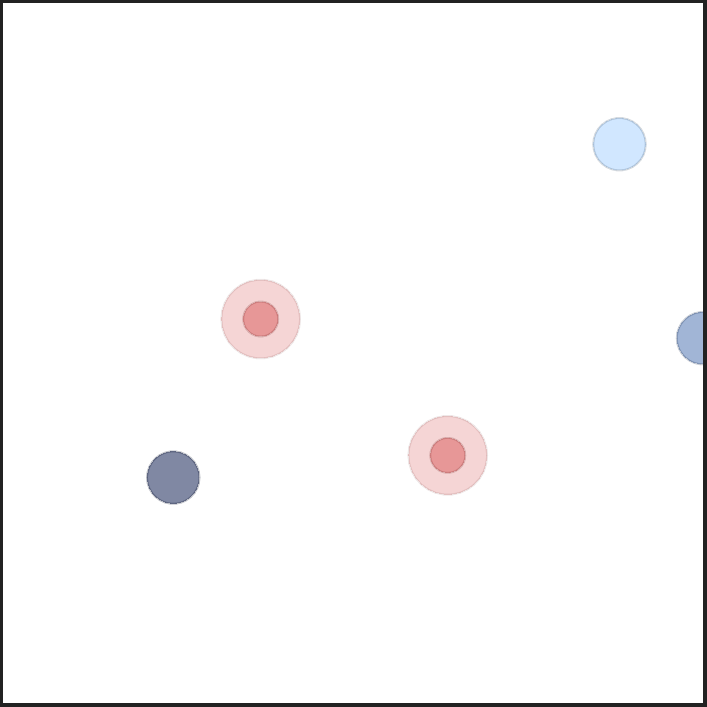}}
\subfigure[] {\includegraphics[height=0.8in,width=0.8in,angle=0]{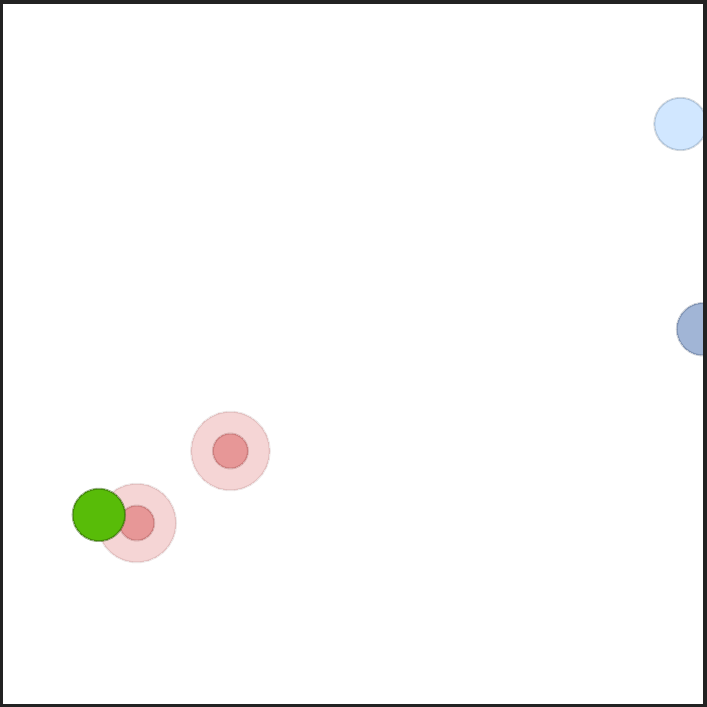}}
\subfigure[] {\includegraphics[height=0.8in,width=0.8in,angle=0]{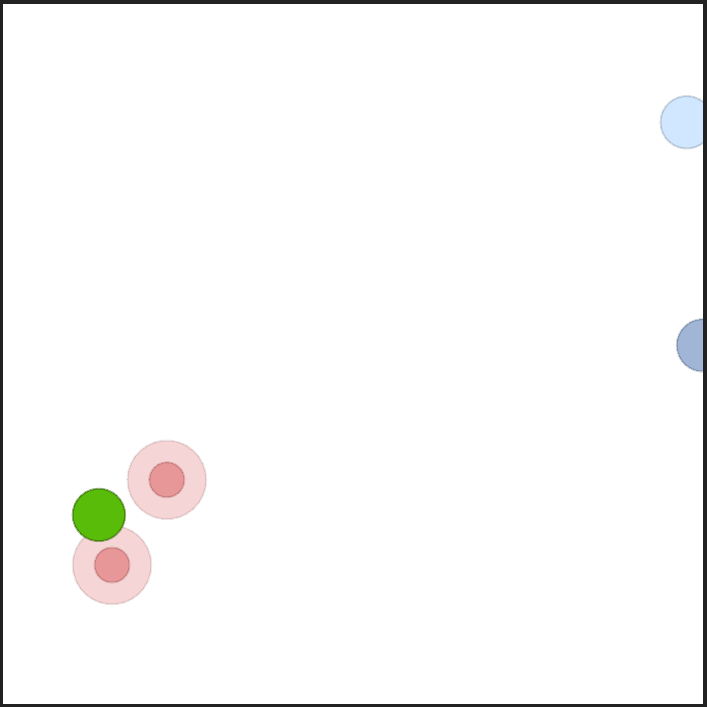}}
\subfigure[] {\includegraphics[height=0.8in,width=0.8in,angle=0]{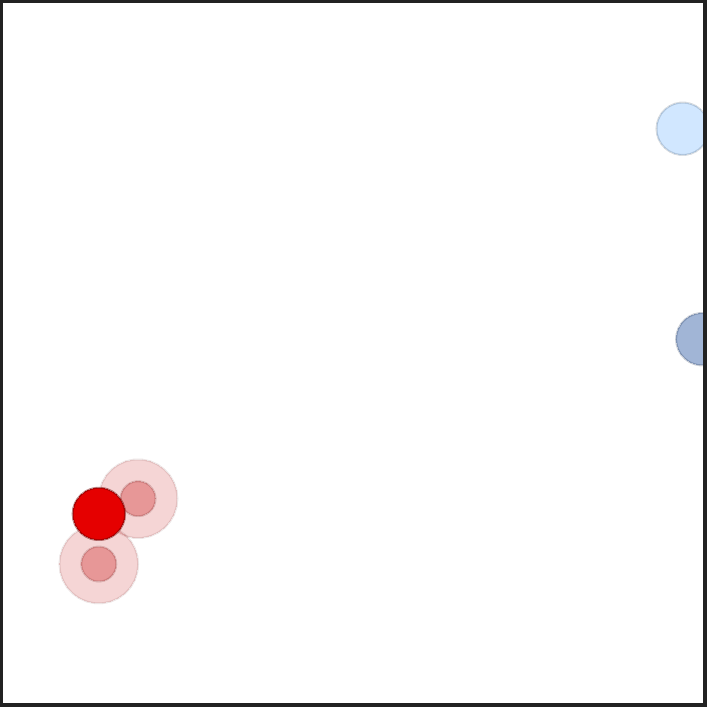}}
\caption{(a) Two rescue agents are in red (translucent red represents the grasper of each agent) and three wounded animals are in blue (deeper blue represents higher value and higher risk); (b), (c) The animal in dark blue turns to green, which means it has been hold by a rescue agent (can't move anymore); (d) The animal in dark blue turns to red, which means it has been captured and saved by the two rescue agents and the episode finished.}
\label{Figure:cooperative-predator-prey}
\end{figure}

\textbf{Game Settings.}
Cooperative endangered wildlife rescue is a more tough version (sparse rewards and high penalty of miss-coordination) of the "predator-prey" task illustrated in MADDPG \cite{lowe2017multi}, which requires more accurate cooperation. There are $N$ slower cooperating rescue agents which cooperatively chase and rescue one of the $M$ faster wounded animals in a randomly generated environment. Each time if all the cooperative rescuers capture 
a wounded animal \textbf{simultaneously}, the agents will be rewarded by some rewards based on the wounded animal they saved. Different wounded animals (e.g., Lion, wildebeest and Deer) correspond to different rewards and different risks. Different risks means that the penalties for miss-coordination on different animals are different (e.g., hurt by the lion). The target for each rescue agent is learning to rescue the same wounded animal independently without knowing each other's policy. Besides, in our settings, we stipulate a rescue agent can hold a wounded animal without suffering any penalty for some game steps $T_{hold}$ before the other partners' arrival. So, the difficulty level of the task can be modulated by the value of $T_{hold}$. The larger, the easier. In the following experiments, we set $N$ to $2$, $M$ to $3$ and $T_{hold}$ to 8. An typical illustration of the cooperative endangered wildlife rescue task is shown in Figure (\ref{Figure:cooperative-predator-prey}).

\textbf{States and Actions.}
 All rescue agents and wounded animals can observe the relative positions of others. Besides, each rescue agent can observe the relative velocities of wounded animals (can't observe the other rescuers' velocities). Actions are accelerations in 4 directions (controlled by 2 actions actually: north or south, east or west). The acceleration of a direction is controlled by the force applied. To sum up, the action space is continuous with a valid range of [-1, 1].

\begin{table}[]
\vspace{-0.3cm}  %
\setlength{\abovecaptionskip}{0.1cm}   
\caption{The payoff matrix of the rescue agents at the terminal state of each episode. Both agents receive the same payoff written in the corresponding cell.}
\begin{tabular}{clcccc}
\textbf{}                                              & \multicolumn{5}{c}{\textbf{Agent 2}}                                                                                                                                                                                                                   \\ \cline{2-6}
\multicolumn{1}{c|}{\multirow{5}{*}{\textbf{Agent 1}}} & \multicolumn{1}{l|}{}                                                       & \multicolumn{1}{l|}{catch a} & \multicolumn{1}{l|}{catch b} & \multicolumn{1}{l|}{catch c} & \multicolumn{1}{c|}{\begin{tabular}[c]{@{}c@{}}on the \\ road\end{tabular}} \\ \cline{2-6}
\multicolumn{1}{c|}{}                                  & \multicolumn{1}{l|}{catch a}                                                & \multicolumn{1}{c|}{11}      & \multicolumn{1}{c|}{-30}     & \multicolumn{1}{c|}{0}       & \multicolumn{1}{c|}{-30}                                                    \\ \cline{2-6}
\multicolumn{1}{c|}{}                                  & \multicolumn{1}{l|}{catch b}                                                & \multicolumn{1}{c|}{-30}     & \multicolumn{1}{c|}{7}       & \multicolumn{1}{c|}{6}       & \multicolumn{1}{c|}{-10}                                                    \\ \cline{2-6}
\multicolumn{1}{c|}{}                                  & \multicolumn{1}{l|}{catch c}                                                & \multicolumn{1}{c|}{0}       & \multicolumn{1}{c|}{6}       & \multicolumn{1}{c|}{5}       & \multicolumn{1}{c|}{0}                                                      \\ \cline{2-6}
\multicolumn{1}{c|}{}                                  & \multicolumn{1}{l|}{\begin{tabular}[c]{@{}l@{}}on the\\  road\end{tabular}} & \multicolumn{1}{c|}{-30}     & \multicolumn{1}{c|}{-10}     & \multicolumn{1}{c|}{0}       & \multicolumn{1}{c|}{0}                                                      \\ \cline{2-6}
\end{tabular}
\label{Figure:extended-climbing-games}
\end{table}

\begin{figure}[!htb]
\vspace{-0.1cm}  %
\setlength{\abovecaptionskip}{0.2cm}   
\setlength{\belowcaptionskip}{-0.3cm}   
\centering
{\includegraphics[height=1.7in,width=2.8in,angle=0]{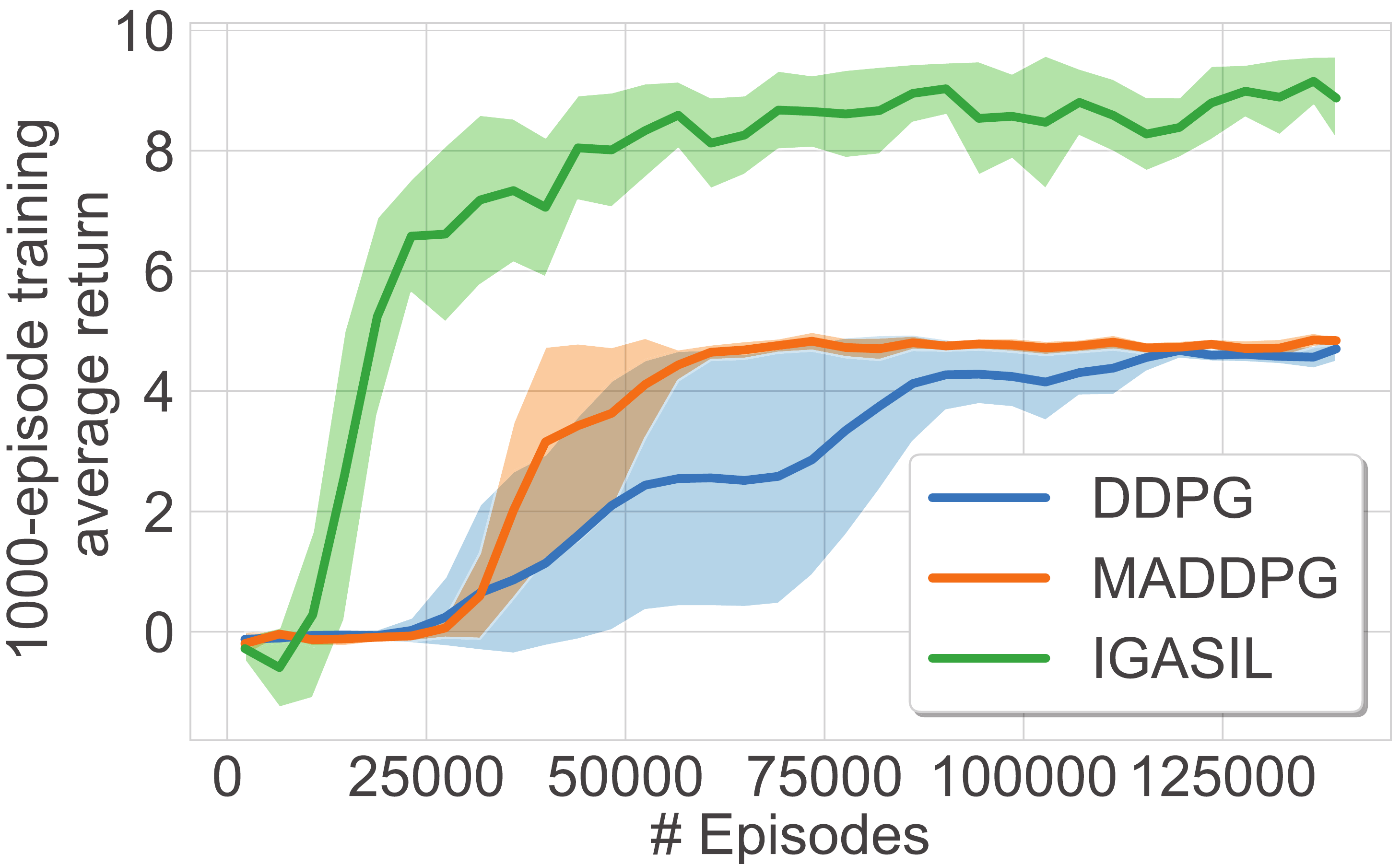}}
\caption{The 1000-episode averaged return of IGASIL versus maddpg and ddpg during training in cooperative endangered wildlife rescue task from the viewpoint of the two rescue agents.}
\label{Figure:exp:predator-prey-rewards}
\end{figure}

\begin{figure*}[!htb]
\vspace{-0.2cm}  %
\setlength{\abovecaptionskip}{0.1cm}   
\setlength{\belowcaptionskip}{-0.2cm}   
\centering
\subfigure[5m] {\includegraphics[height=1.9in,width=3.2in,angle=0]{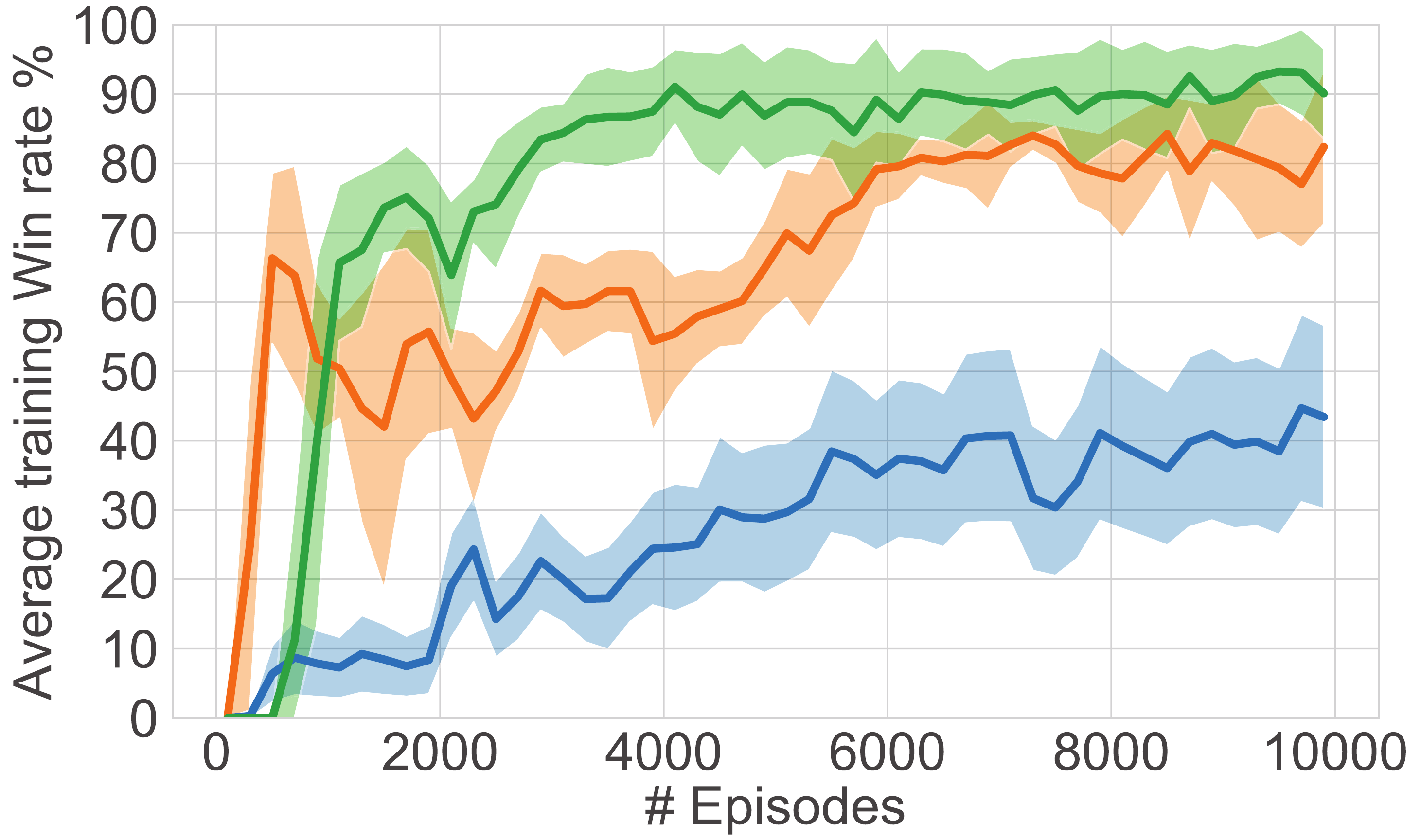}}
\subfigure[2d3z] {\includegraphics[height=2.1in,width=3.2in,angle=0]{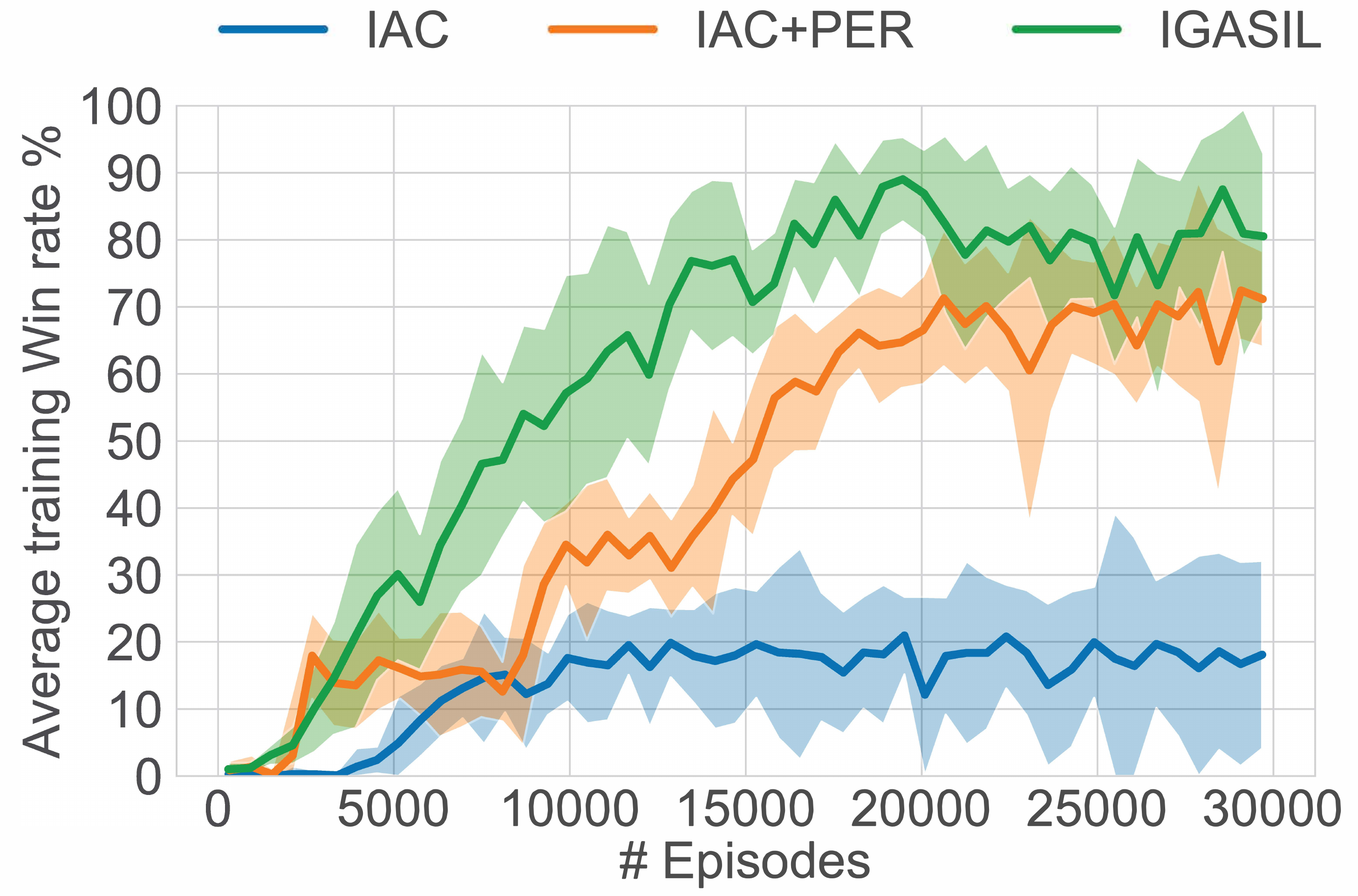}}
\caption{Win rates for IGASIL, IAC and IAC+PER on two different scenarios.}
\label{Figure:exp:starcraft-learning-curves}
\end{figure*}

\textbf{Reward Function.}
The environmental rewards are sparse and only depend on the terminal state of each episode. The payoff matrix of the terminal state is defined in Table (\ref{Figure:extended-climbing-games}). At each state $s_T$, if both agents capture and save the same target $a, b $ or $c$ simultaneously, they will both receive $11, 7 $ or $5$. Else, if one agent captures $i$ while the other captures $j$, they will both receive $r(i,j)$. Finally, if one agent holds $i$ and the other doesn't come over in $T_{hold}$ steps, they will both be punished by $r(i,3)$. According to Table (\ref{Figure:extended-climbing-games}), the theoretical-optimal action is "\emph{catch a}", but the action "\emph{catch c}" can be easily mistaken for having the highest reward due to the lowest penalty for exploration. The game can be seen as a Markov extension of the climbing game \cite{claus1998dynamics} with continuous state and action spaces.

\begin{figure}[!htb]
\setlength{\abovecaptionskip}{0.1cm}   
\setlength{\belowcaptionskip}{-0.5cm}   
\centering
{\includegraphics[height=1.7in,width=2.8in,angle=0]{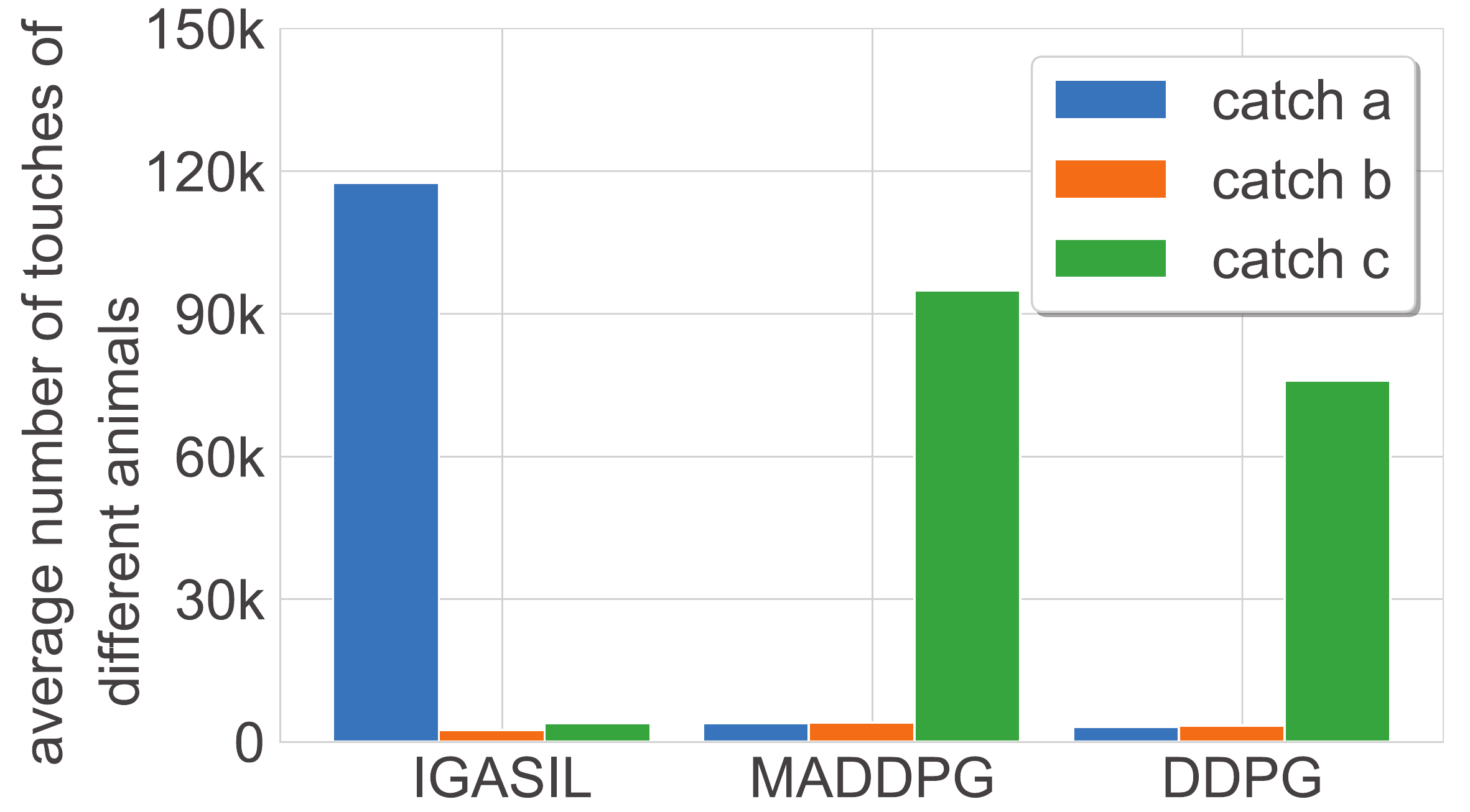}}
\caption{The average number of touches of different wounded animals by the two rescues during training with different algorithms.}
\label{Figure:action_selection_times}
\end{figure}

\textbf{Experimental Results}
To make the different algorithms comparable, we pre-train both the rescue agents and wounded animal agents with DDPG and save the animal models during training. Then, we reuse the same pre-trained animal models as the default policies for the wounded animals in all experiments. The learning curves of IGASIL versus MADDPG and DDPG are plotted in Figure (\ref{Figure:exp:predator-prey-rewards}) under five random seeds (1,2,3,4,5). To show a smoother learning procedure, the reward value is averaged every 1000 episodes. Apparently, our IGASIL outperforms MADDPG and DDPG by a significant margin in terms of both convergence rate and final performance. To obviously express the different equilibrium the three algorithms converged to, we show the average number of touches of the three wounded animals by the two rescue agents during training with different algorithms in Figure (\ref{Figure:action_selection_times}). As illustrated in Table (\ref{Figure:extended-climbing-games}), "animal a" has the highest reward +11 and the highest miss-coordination penalty -30. In Figure (\ref{Figure:action_selection_times}), we can easily observe that only our IGASIL succeeds in converging to the optimal Nash equilibrium ($a,a$) (learned to rescue "animal a") while MADDPG and DDPG converge to the worst equilibrium ($c,c$) (learned to rescue "animal c"). This result shows that only our IGASIL overcame the risk of being punished by miss-coordination and achieved a better cooperation result (which need more accurate collaborations).

\subsubsection{Decentralised StarCraft Micromanagement}

\

\textbf{Game Settings.}
In this section, we focus on the problem of micromanagement in StarCraft, which refers to the low-level control of individual units' positioning and attack commands as they fight with enemies. This task is naturally represented as a multiagent system, where each StarCraft unit is controlled by a decentralized independent controller (agent). We consider two scenarios with symmetric teams formed of: 5 marines (5m) and 2 dragoons with 3 zealots (2d\_3z). The enemy team is controlled by the built-in StarCraft AI, which uses reasonable but suboptimal hand-crafted heuristics. Since the game is easily obtained and is fair for comparison, micromanagement of StarCraft has become a standard testbed for multagent learning algorithms (for both independent learners and joint learners), which has been widely studied in recent years such as COMA \cite{foerster2017counterfactual}, BiCNet \cite{peng2017multiagent}, QMIX \cite{rashid2018qmix}. Different from their approaches which are all joint learners, we focus on fully independent learning paradigm (independent learning and independent execution). Similar settings can be found in \cite{foerster2017stabilising}. The settings of action space, state features and reward function are similar to that in COMA \cite{foerster2017counterfactual}. 

\textbf{Experimental Results.} Figure (\ref{Figure:exp:starcraft-learning-curves}) shows the average training win rates as a function of episode for each method and for each StarCraft scenario. For each method, we conduct 5 independent trials and calculate the win rate every 200 training episodes and average them across all trials. 
In Figure (\ref{Figure:exp:starcraft-learning-curves}), IAC represents the independent Actor-Critic. The actor and critic parameters are also shared among the homogeneous agents. IAC+PER 
 means we add a positive replay buffer $M_E^i$ to each IAC but each agent only stores the original entire trajectory into $M_E^i$ instead of additionally sampling and storing some sub-trajectories (sub-skills). 
IGASIL is our approach, which equals to IAC+SCER.

\begin{table}[h]
\vspace{-0.1cm}  
\setlength{\abovecaptionskip}{3pt}%
\setlength{\belowcaptionskip}{0pt}
\caption{Mean win percentage across final 1000 evaluation episodes for the different scenarios. The highest mean performances are in bold. The results of COMA are extracted from the published paper \cite{foerster2017counterfactual}.}
\centering
\setlength{\tabcolsep}{2.9mm}{
\begin{tabular}{|c|c|c|c|c|c|c|}
  \hline
  Map & Heur. & IAC & IAC+PER & COMA & IGASIL\\
  \hline
  5 M & 66 & 45 & 85 & 81 & \textbf{96} \\
  \hline
  2d3z & 63 & 23 & 76 & 47 & \textbf{87} \\
  \hline
\end{tabular}
}
\label{Table:evaluation}
\end{table}

The results show that our IGASIL is superior to the IAC baselines in all scenarios. For the parameters sharing among the homogeneous agents (which has been shown to be useful in facilitate training \cite{gupta2017cooperative}), IAC also learned some coordination on the simpler m5v5 scenario. However, it's hard for IAC to achieve cooperation on the more complicated 2d3z scenario, due to the different types of units, local observations and the resulting dynamics. On 2d3z, our IGASIL still achieves a 82\% win rate during training and achieves a 87\% win rate in evaluation. 
In Table (\ref{Table:evaluation}), we summarize the averaged evaluation win rates of different approaches under multiple combat scenarios (The results of COMA are extracted from the published paper \cite{foerster2017counterfactual}.). The winning rate of the built-in AI (Heur.) is also provided as an indicator of the level of difficulty of the combats. The best result for a given map is in bold. The results show that our independent IGASIL outperforms all approaches in the performance of evaluation win rate and even outperforms the centralized trained COMA. Besides, our IGASIL converges faster than COMA according to  Figure (\ref{Figure:exp:starcraft-learning-curves}) and Figure (3) of  \cite{foerster2017counterfactual}.

\subsection{Sample Efficiency of IGASIL} \label{exp:sample-efficient}
\begin{figure}[!htb]
\setlength{\abovecaptionskip}{0.2cm}
\centering
{\includegraphics[height=1.4in,width=3in,angle=0]{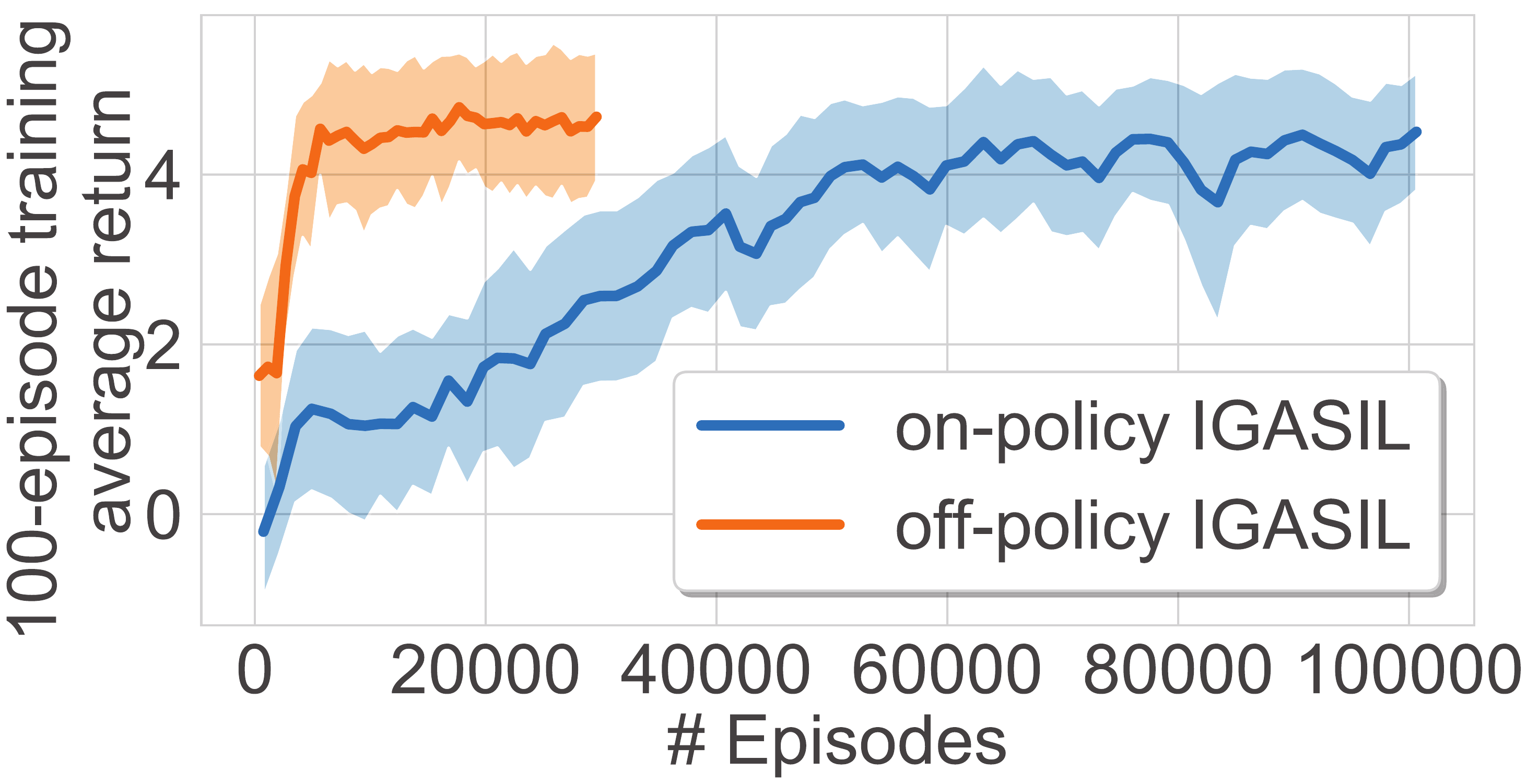}}
\caption{The 100-episode training average returns of on-policy and off-policy IGASIL respectively.}
\label{Figure:exp:on-off-policy}
\end{figure}

To show the sample efficiency of our off-policy IGASIL in cooperative multiagent systems, we compare the performance of the on-policy (on-policy AC+SCER) and off-policy versions (based on off-policy AC+SCER) of IGASIL in the animal rescue task. To clearly see the influence of the on-policy and off-policy only, we initialize $M_E^i$ for each agent $i$ with the same 32 demonstrations (demonstrated by the pre-trained DDPG agent in Section \ref{exp:effectiveness:predator-prey}, which will cooperatively catch "animal c", resulting an average return +5). We perform learning under 5 random seeds (1,2,3,4,5). The imitation learning results are shown in Figure (\ref{Figure:exp:on-off-policy}). The x-axis represents the number of episodes interacting with the environment during training. From the figure, we see both the on-policy and off-policy IGASILs finally achieved cooperation (learned to "catch c" simultaneously). However, the off-policy IGASIL is able to recover the expert policy given a significantly smaller number of samples than the on-policy version (about 10 times less). Thus, the sample-efficiency was significantly improved.

\begin{figure}[!htb]
\setlength{\abovecaptionskip}{0.2cm}
\setlength{\belowcaptionskip}{-0.6cm}
\centering
{\includegraphics[height=1.6in,width=2.8in,angle=0]{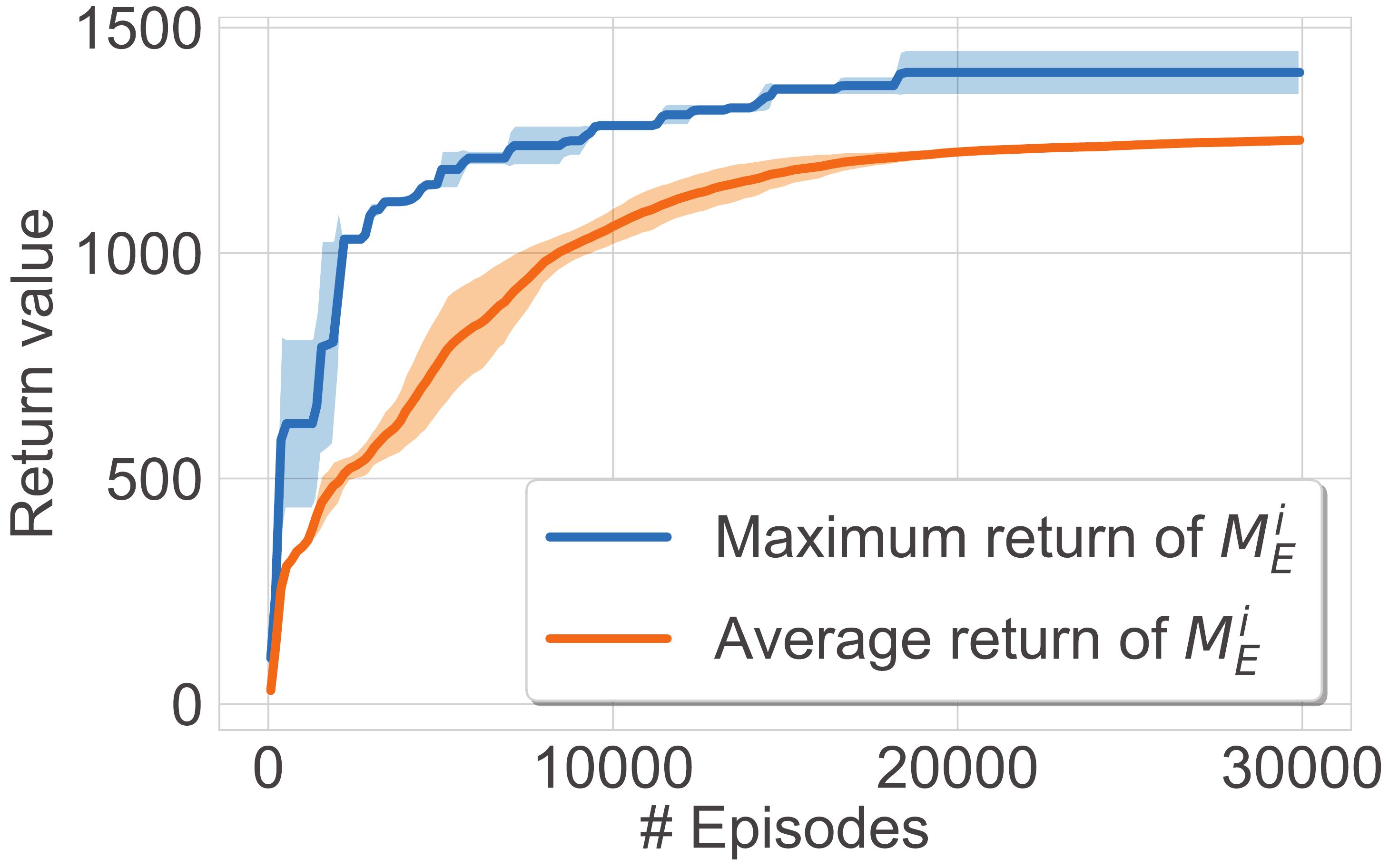}}
\caption{The curves of the maximum and average return values of the trajectories/sub-trajectories stored in $M_E^i$ during training on the 2d3z scenario.}
\label{Figure:exp:positive-return}
\end{figure}

\begin{figure}[!htb]
\setlength{\abovecaptionskip}{0.2cm}
\setlength{\belowcaptionskip}{-0.2cm}
\centering
{\includegraphics[height=1.5in,width=2.7in,angle=0]{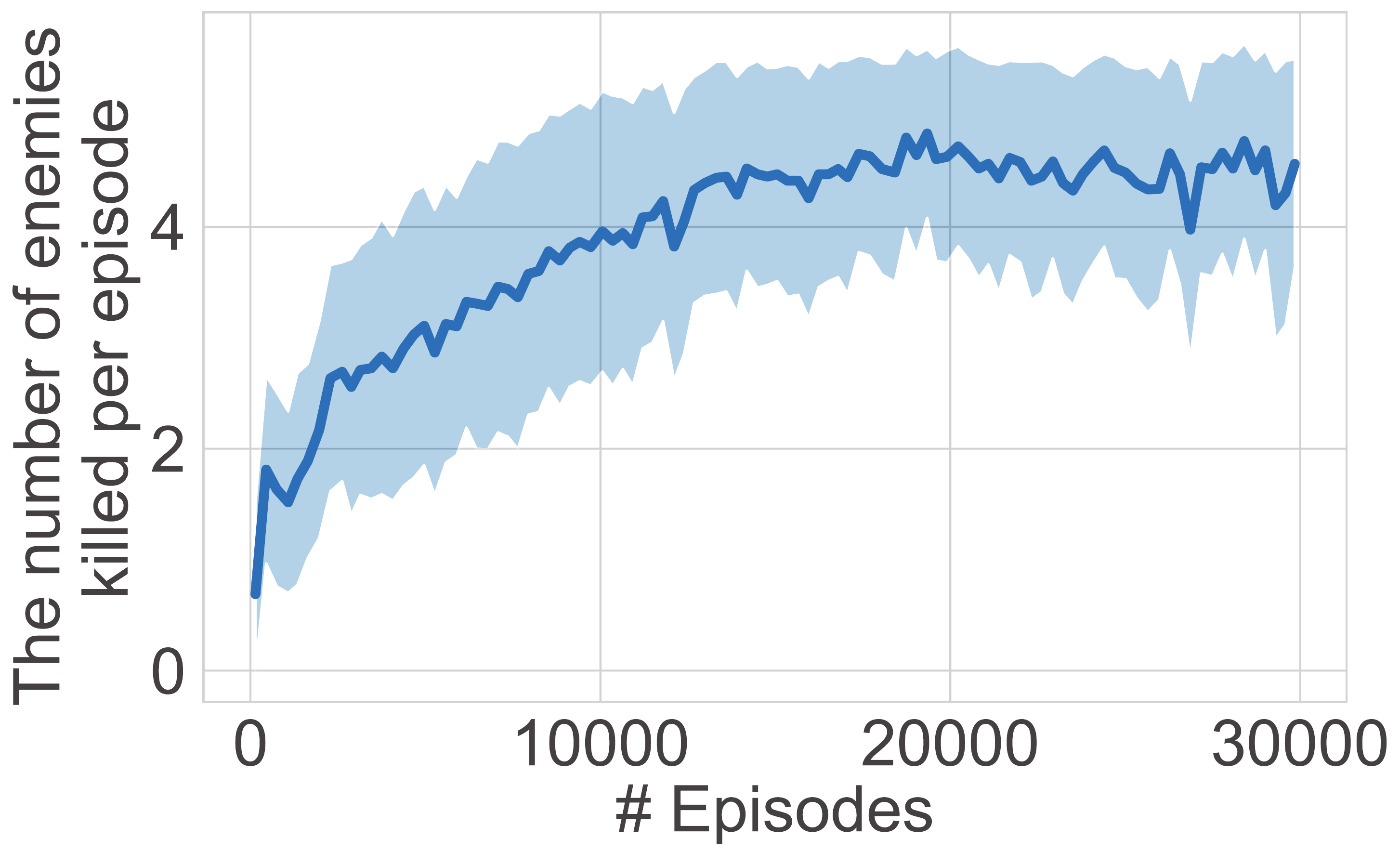}}
\caption{The curve of the number of enemies killed per episode during training on the 2d3z scenario (5 enemies in total).}
\label{Figure:exp:killed enemies}
\end{figure}

\subsection{Contributions of Sub-Curriculum Experience Replay} \label{exp:effect-PHER}
We analyze the roles and contributions of the sub-curriculum experience replay here. In Figure (\ref{Figure:exp:starcraft-learning-curves}), we see IAC+PER outperforms the IAC baseline in both scenarios, which means adding a positive buffer $M_E^i$ to store the good trajectories and doing self-imitation learning could help the independent agents reach a cooperation. Besides, IGASIL (IAC+SCER) outperforms IAC+PER in both scenarios, which indicates that adding additional sub-trajectories sampled from the original one (useful sub-skills) to $M_E^i$ could further accelerate the self-imitation procedure and achieve  better cooperation. Besides, from Figure (\ref{Figure:exp:positive-return}), we see that as the self-imitation learning progresses, better experiences are stored in $M_E^i$. Then, the agents would grasp these beneficial skills stored in $M_E^i$ via self-imitation learning. Using the learned new skills, the agents are more likely to reach a better cooperation and learn better policies. As shown in Figure (\ref{Figure:exp:killed enemies}), as the stored experiences in $M_E^i$ gets better and better, the number of enemies killed per episode (which indicates the performance of the current policies) grows. All these analysis illustrates that our sub-curriculum experience replay does help the independent agents to achieve better cooperation. However, in our settings, we limit the size of the normal buffers and positive buffers. The techniques proposed in \cite{foerster2017stabilising} might be incorporated into our framework to allow us to use larger buffer size and further improve the sample efficiency. We will consider it as future work.

\section{Conclusion \& Future Work}
In this paper, we presented a novel framework called independent generative adversarial self-imitation learning (IGASIL) to address the coordination problems in some difficult fully cooperative Markov Games. Combining self-imitation learning with generative adversarial imitation learning, IGASIL address the challenges (e.g., non-stationary and exploration-exploitation) by guiding all agents to frequently explore more around the nearby regions of the past good experiences and learn better policy. Besides, we put forward a sub-curriculum experience replay mechanism to accelerate the self-imitation learning process. Evaluations conducted in the testbed of StarCraft unit micromanagement and cooperative endangered wildlife rescue show that our IGASIL produces state-of-the-art results in terms of both convergence speed and final performance. In order to obviously see whether our IGASIL can significantly facilitate the coordination procedure of the interactive agents, we only consider the environments with deterministic reward functions at present. In our future work, we are going to deal with more challenging cooperative tasks (e.g., environments with stochastic rewards).

%
%
\begin{acks}
The work is supported by the National Natural Science Foundation of China (Grant Nos.: 61702362, U1836214), Special Program of Artificial Intelligence, Tianjin Research Program of Application Foundation and Advanced Technology (No.: 16JCQNJC00100), and Special Program of Artificial Intelligence of Tianjin Municipal Science and Technology Commission (No.: 569 17ZXRGGX00150).
%
%
\end{acks}